\documentclass[pra,twocolumn,showpacs,preprintnumbers,amsmath,amssymb,tightenlines,superscriptaddress]{revtex4}

\usepackage{graphicx}
\usepackage{dcolumn}
\usepackage{bm}
\usepackage{epsfig}
\usepackage{stmaryrd}

\newcommand{\expec}[1]{\langle #1 \rangle}

\newcommand{\sub}[2]{{#1}_{\mbox{\!\! \scriptsize #2}}}

\def\beq{\begin{equation}}
\def\eeq{\end{equation}}

\def\CR{\nonumber\\[0.15cm]}
\newcommand{\rref}[1]{Ref.~\cite{#1}}
\newcommand{\fref}[1]{Fig.~\ref{#1}}
\newcommand{\frefp}[2]{Fig.~\ref{#1}~(#2)}

\newcommand{\eref}[1]{Eq.~(\ref{#1})}

\newcommand{\sref}[1]{section~\ref{#1}}

\newcommand{\cref}[1]{chapter~\ref{#1}}

\newcommand{\Cref}[1]{Chapter~\ref{#1}}

\newcommand{\aref}[1]{appendix~\ref{#1}}
\newcommand{\bref}[1]{(\ref{#1})}

\def\hamx{ \hat{H}_{0}(\mathbf{x})}

 \newcommand{\intas}{\int \!\! d^{3} \mathbf{x}\:}
 \newcommand{\intass}{\int \!\! d^{3}  \mathbf{x'}\:}
 
\def\psix{\hat{\Psi}(\mathbf{x})}

\def\psiadx{\hat{\Psi}^{\dagger}(\mathbf{x})}

\def\chix{\hat{\chi}(\mathbf{x})}

\def\phix{\phi(\mathbf{x})}

\def\xv{\mathbf{x}}
\def\xdv{\mathbf{x'}}

\begin{document}
\title{Quantum-field dynamics of expanding and contracting Bose-Einstein condensates}
\author{S.~W\"uster}
\altaffiliation[Present address:
]{School of Physical Sciences, University of Queensland, Brisbane QLD 4072, Australia}
\author{B.~J.~D\c{a}browska-W\"uster}
\altaffiliation[Present address:
]{School of Physical Sciences, University of Queensland, Brisbane QLD 4072, Australia}

\author{S.~M.~Scott}
\author{J.~D.~Close}
\author{C.~M.~Savage}

\affiliation{Department of Physics, Australian National University, Canberra ACT 0200, Australia}
\email{craig.savage@anu.edu.au}

\begin{abstract}
We analyze the dynamics of quantum statistics in a harmonically trapped Bose-Einstein condensate, whose two-body interaction strength is controlled via a Feshbach resonance. From an initially non-interacting coherent state, the quantum field undergoes Kerr squeezing, which can be qualitatively described with a single mode model. To render the effect experimentally accessible, we propose a homodyne scheme, based on two hyperfine components, which converts the quadrature squeezing into number squeezing. The scheme is numerically demonstrated using a two-component Hartree-Fock-Bogoliubov formalism. 
\end{abstract}


\maketitle
\section{Introduction}

Multimode quantum fields are the appropriate description for a vast array of phenomena in high-energy physics, condensed matter physics and cosmology, but they are notoriously difficult to analyze theoretically. Unlike many quantum fields, it appears feasible that those describing degenerate Bose and Fermi gases can be experimentally manipulated, detected and studied. The development of techniques for this could have broad influence on a variety of outstanding problems in physics. 

Understanding the many-body quantum state of a Bose-Einstein condensate (BEC) is relevant for systems including: squeezed atom-lasers \cite{matthias:simon:kerr}, simulations of cosmological particle production in the early universe \cite{visser:review,calz:hu,calz:hu2, barcelo:cpc}, and the quantitative description of collapsing condensates due to attractive interactions \cite{savage:coll,wuester:nova, wuester:nova2}. 

The quantum field equations describing expanding and collapsing condensates are analogues of those describing a quantum field on a curved space-time \cite{visser:review}. Learning to experimentally manipulate and study the analogue system provided by expanding and collapsing condensates promises us a new window on processes such as signature change that may be relevant for the evolution of the early universe \cite{silke:sigchange}. The quantum field of expanding and contracting condensates is the subject of this paper. 

While quantum-field models exist to approximately describe the time-evolution of Gaussian quantum states in such non-equilibrium situations \cite{griffin:gappy,hutchinson:gaplessmodes,morgan:thermalqft,holland:burst,steel:wigner,book:qn,norrie:prl,castin:validity, matthewandblair:pgpe,norrie:thesis}, they require us to specify the initial quantum state of the condensate, which is complicated if interactions are present \cite{dunningham:sqz,lewenstein:phasediff,haque:squeezing}.

Due to the controllability of atomic interactions by Feshbach resonances, quantum field dynamics in a BEC can however be examined starting from a non-interacting initial state. Then, we assume that the many-body quantum state is represented by a coherent state; perhaps originating from a mixture of all different phases, as in the optical laser \cite{molmer:cohstate}. This initial situation was realized in experiments on collapsing Bose-Einstein condensates with attractive interactions \cite{jila:nova}. We also theoretically consider it here, but with interactions suddenly rendered attractive \emph{or} repulsive.

We show that the condensate's state evolves from coherent to quadrature squeezed due to the Kerr effect. For repulsive interactions a single-mode model provides a qualitative description of the squeezing. However, a multi-mode analysis is required for a quantitative description, especially in the attractive case.

The experimental detection of quadrature squeezing requires a phase-reference, such as in homodyne detection \cite{book:walls:milburn}. We propose such a scheme based on a splitting of the Bose-Einstein condensate into two separately conserved hyperfine components. We demonstrate the proposal using two-component Hartree-Fock-Bogoliubov (HFB) theory with realistic experimental parameters. The homodyne detection is found reliable even in the presence of imperfections like interactions between the local oscillator component and the squeezing field. Also the Kerr squeezing of the local oscillator itself does not prevent us from attaining a measurable reduction of the number variance.

A single mode model of Kerr squeezing has previously been found useful despite the presence of multi-mode effects \cite{matthias:simon:kerr}. The authors of \rref{matthias:simon:kerr} also observe that the interference of two quadrature squeezed atom-lasers can yield a number squeezed state, in accordance with our successful simulation of the homodyne scheme with squeezed local oscillator. In contrast to \rref{matthias:simon:kerr}, which is focussed on the creation of a squeezed atom laser, we consider a simpler setup and target studies of quantum field dynamics. A core ingredient in our work is a Feshbach resonance, allowing the use of a simple coherent initial state.

This paper is organized as follows. Section \ref{methods} provides a brief overview of the three quantum theories employed: the single mode model, HFB theory and the truncated Wigner approximation (TWA). In \sref{groundstate} we analyse Kerr squeezing in a harmonically trapped condensate and in \sref{homodyne} propose a matter-wave homodyne scheme to detect it. Technical details regarding variances in the HFB formalism as well as the two-component HFB equations of motion can be found in the appendix.

\section{Methods \label{methods}}

An ensemble of Bose condensed atoms in a harmonic trap is described by the many-body Hamiltonian
\begin{align}
\hat{H} =& \intas \psiadx \hamx \psix
\CR
&
 + \frac{\sub{U}{0}}{2}  \psiadx  \psiadx  \psix \psix,
\label{Hamiltonian}
\end{align}
where
\begin{align}
\hamx= -\frac{\hbar^2}{2m} \nabla_\mathbf{x}^2 +V(\mathbf{x})
\label{singleparthamil}
\end{align}
is the single particle Hamiltonian, and $\hat{\Psi}(\mathbf{x})$ denotes the field operator in the Heisenberg picture that annihilates atoms of mass $m$ at position $\mathbf{x}$. We have assumed a contact interaction of strength $\sub{U}{0}=4\pi\hbar^{2}a_{s}(t)/m$ with time dependent scattering length $a_{s}(t)$ and a spherically symmetric harmonic potential $V(\mathbf{x})=m\omega^{2}\mathbf{x}^{2}/2$. Note that we use the physical coupling for the parameter $\sub{U}{0}$ directly, rather than the bare coupling, which we justify in \sref{hfb}.
The Heisenberg equation for the field operator is:
\begin{align}
i\hbar\frac{\partial  \psix}{\partial t} &= \hat{H}_{0} \psix + \sub{U}{0} \psiadx  \psix \psix.
\label{Heisenberg_atoms}
\end{align}
In the following subsections we briefly introduce several methods to obtain approximate solutions to this multi-mode quantum field problem.

\subsection{Single mode Kerr squeezing \label{oscimodes}}

Among the single particle bases in which we can analyze the atom-field dynamics of \eref{Heisenberg_atoms}, the harmonic oscillator basis takes a special role. The condensate is initially assumed to be non-interacting and in the trap ground state. Thus in the oscillatory basis we assume it is in a coherent state of one single particle mode.

Let us expand the field operator as:
$
\psix=\sum_{k=0}^{\infty}\varphi_{k}(\xv)\hat{a}_{k}.
$
The $\varphi_{k}(\xv)$ are the eigenstates of the single particle Hamiltonian for a harmonic potential, with a collective index $k$ labelling all quantum numbers. The operator $\hat{a}_{k}$ annihilates an atom in eigenstate $\varphi_{k}(\xv)$, with 
$\hat{H}_{0}\varphi_{k}(\xv)=\hbar \omega_{k}\varphi_{k}(\xv)$. Using this expansion, we can rewrite \eref{Heisenberg_atoms} as
\begin{align}
i\hbar\frac{\partial}{\partial t}  \hat{a}_{k}&= \hbar \omega_{k} \hat{a}_{k}+ \sum_{lmn} {U}_{klmn} \hat{a}^{\dagger}_{l} \hat{a}_{m} \hat{a}_{n}.
\label{osci_gpe}
\end{align}
The ${U}_{klmn}$ are overlap integrals of the form
\begin{align}
{U}_{klmn}&=\sub{U}{0}\intas \varphi_{k}^{*}(\xv) \varphi_{l}^{*}(\xv) \varphi_{m}(\xv) \varphi_{n}(\xv).
\label{Uklmns}
\end{align}
Initially all the atoms are in state $\varphi_{0}(\xv)$, and for short times we can approximate \eref{osci_gpe} by:
\begin{align}
i\hbar\frac{\partial}{\partial t}  \hat{a}_{0}&= \hbar \omega_{0} \hat{a}_{0}+ {U}_{0000} \hat{a}^{\dagger}_{0} \hat{a}_{0} \hat{a}_{0}.
\label{singlemode_gpe}
\end{align}
Using the trap ground state $\varphi_{0}(\xv)=A \exp{[-r^{2}/(2 \sigma^{2})]}$ with $A=(\pi \sigma^{2})^{-3/4}$, $\sigma=(\hbar m/\omega)^{-1/2}$, we find
$
\sub{U}{0000}=\sub{U}{0} (2\pi\sigma^{2})^{-3/2}.
$
The ground state energy term in \eref{singlemode_gpe} can be eliminated using rotating frame operators: $\hat{b}=\exp{(i \omega_{0} t)}\hat{a}_{0}$. The equation of motion then becomes:
$
i\frac{\partial}{\partial t}  \hat{b}= -\chi \hat{b}^{\dagger} \hat{b} \hat{b},
$
with $\chi=\sub{U}{0000}/\hbar$.
It is known that this operator equation gives rise to Kerr squeezing \cite{book:walls:milburn} 
in the evolution of the quantum state. Further details can be found in \rref{matthias:simon:kerr}. Here we merely state the most important facts. 

One can define quadratures for the state $\varphi_{0}(\xv)$ as:
\begin{align}
\hat{X}^{\theta}&=\hat{a}_{0}^{\dagger}e^{i\theta}  +\hat{a}_{0} e^{-i\theta}.
\label{thetaquadratops}
\end{align}
The variance of these operators, $[\Delta \hat{X}^{\theta}]^2=\expec{[\hat{X}^{\theta}]^{2}} -\expec{\hat{X}^{\theta}}^{2}$, gives information about the shape of the quantum state's Wigner functon in phase-space \cite{book:walls:milburn}. In the Kerr effect the variance in one quadrature $\sub{\theta}{sqz}$ drops below the value for a coherent state ($[ \Delta\hat{X}^{\theta}]^{2}=1$), while for the orthogonal quadrature $\sub{\theta}{sqz}+\pi/2$ it increases.

\subsection{Hartree-Fock-Bogoliubov theory\label{hfb}}

To go beyond the previous section and include multi-mode aspects of the quantum evolution, we make use of the HFB formalism \cite{griffin:gappy,hutchinson:gaplessmodes,morgan:thermalqft, holland:burst}. Thus we decompose $\psix$ into a condensate part $\phix$ and quantum fluctuations $\chix$, such that $\psix = \phix + \chix$ and $\langle \psix \rangle= \phix$. The quantum fluctuations can be described in terms of their lowest order correlation functions: the normal density $G_{N}(\mathbf{x},\mathbf{x}') = \langle \hat{\chi}^{\dagger} (\mathbf{x}')\hat{\chi}(\mathbf{x}) \rangle$ and anomalous density $G_{A}(\mathbf{x},\mathbf{x}') = \langle \hat{\chi}(\mathbf{x}') \hat{\chi}(\mathbf{x}) \rangle$. The resulting equations of motion and their implementation for a spherically symmetric, trapped condensate have been described in Refs.~\cite{wuester:nova,wuester:nova2}. 

We now explain how to calculate variances of the oscillator ground state quadratures \bref{thetaquadratops} in HFB theory. Since we have
$
\hat{a}_{k}=\intas \varphi_{k}^{*}(\xv)\hat{\Psi}(\xv),
$
we can use
\begin{align}
\expec{\hat{a}^{\dagger}_{k}\hat{a}_{k'}}&=\intas \intass 
\varphi_{k'}(\xdv) \varphi_{k}^{*}(\xv)\expec{\hat{\Psi}^{\dagger}(\xdv)\hat{\Psi}(\xv)}
\label{oscicorrfct}
\end{align}
Expressions like \eref{oscicorrfct} are all we need to extract the quadrature variance of the trap ground state $[\Delta \hat{X}^{\theta}]^{2}$ from our HFB simulations using spherical symmetry (described in \cite{wuester:nova}). The result for 
the variance of the $\theta$ quadrature is:
\begin{align}
[\Delta \hat{X}_{0}^{\theta}]^{2}&=1 + 2 \intas \intass \bigg[\varphi_{0}(\xdv)\varphi_{0}(\xv)^{*}
G_{N}(\xv,\xdv)
\CR 
&+2\mathfrak{Re}\left\{\varphi_{0}^{*}(\xdv)\varphi_{0}^{*}(\xv)
G_{A}(\xv,\xdv) e^{-2i\theta}\right\} \bigg].
\label{quadrat_variance_groundst}
\end{align}

For our analysis of the homodyne scheme we are also interested in the variance of the total atom number.
Using $\hat{N}=\intas\hat{\Psi}^{\dagger}(\xv)\hat{\Psi}(\xv)$ we derive:
\begin{align}
&[\Delta \hat{N}]^{2}=\sub{N}{tot} + \intas\intass  \bigg[
\CR
&2 \mathfrak{Re}\left\{\phi^{*}(\xv)\phi^{*}(\xdv) G_{A}(\xv,\xdv)  \right\}
+ 2\phi^{*}(\xv)\phi(\xdv) G_{N}(\xv,\xdv) 
\CR
&+\left[\left|G_{A}(\xv,\xdv) \right|^{2} + \left|G_{N}(\xv,\xdv) \right|^{2}  \right] \bigg]. 
\label{single_comp_numbvervar1}
\end{align}
Further details about the form of Eqs.~\bref{quadrat_variance_groundst} and \bref{single_comp_numbvervar1} that we use for numerical solutions in spherically symmetric situations, are given in \aref{hfb_variances_appendix}. 

In this work we have found that the results are independent of the numerical cutoff $K=\pi/\Delta x$, where $\Delta x$ is the grid spacing, only \emph{without} coupling renormalization. As previously noted \cite{holland:deltamu,morgan:thermalqft}, the diagonal part of $G_{A}$ is the only divergent quantity in the formalism. However, this contributes negligibly to the variances of interest here,  Eqs.~\bref{quadrat_variance_groundst} and \bref{single_comp_numbvervar1}.

\subsection{Truncated Wigner theory\label{twa}}

We have highlighted the value of verifying numerical quantum-field theory results for a BEC by using two quite different formalisms in \rref{wuester:nova2}. Here we follow the same approach, by investigating quadrature squeezing using the HFB method as well as the truncated Wigner approximation (TWA) \cite{steel:wigner,book:qn,norrie:prl,castin:validity,norrie:thesis}. We have given a compact summary of the method and its implementation in \rref{wuester:nova2}. Our TWA simulations are based on solutions of the relevant stochastic differential equation in the harmonic oscillator basis \cite{matthewandblair:pgpe}. The theory is then expressed in terms of the stochastic wave function $\alpha(\xv)=\sum_{n}\alpha_{n}\varphi_{n}(\xv)$.

To determine the quadrature variances in this framework, we use the appropriate correspondences between averages of the stochastic wavefuntions and operator expectation values. 
Most importantly \cite{book:qn}:
\begin{align}
\overline{\alpha^{*}_n \alpha_m}&\rightarrow
\frac{1}{2} \left(\expec{\hat{\Psi}^{\dagger}_n \hat{\Psi}_m} + \expec{\hat{\Psi}_m\hat{\Psi}^{\dagger}_n} \right)
\label{op_corresp}
\end{align}
Using \eref{op_corresp} we obtain
\begin{align}
[\Delta \hat{X}^{\theta}_{0}]^{2}=&2\left(\overline{\alpha^{*}_{0}\alpha_{0}} - | \overline{\alpha_{0}}|^{2} \right)
\CR
 &+2\mathfrak{Re}\left\{ \left(\overline{\alpha_{0}\alpha_{0}} -  \overline{\alpha_{0}}^{2} \right)e^{-2i\theta} \right\}.
\label{quadrat_uncert_twa}
\end{align}
%

\section{Squeezing of the Ground-State \label{groundstate}}

In this section we discuss our simulations of ground state quadrature squeezing in a harmonically trapped condensate using the TWA and HFB. We begin with a $^{87}$Rb condensate of $6000$ atoms in a spherical trap with $\omega=12.8 \times 2 \pi$~Hz. Initially, the interaction between the atoms is tuned to zero by use of a Feshbach resonance so that all atoms occupy the trap ground state. The scattering length is then suddenly switched to an either attractive or repulsive value, which we label $\sub{a}{dyn}$.

\begin{figure}
\centering
\epsfig{file={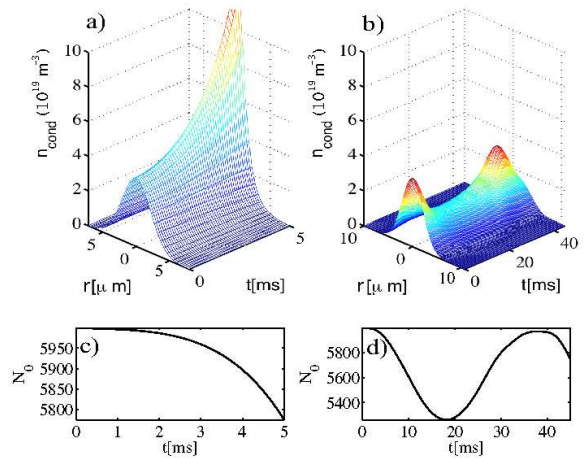},width=\columnwidth} 
\caption{(Color online) Evolution of the condensate density and ground state population for scenarios I  (a,c) and II (b,d), as described in the text. The ground state population in panels (c) and (d) is defined by $N_{0}=\intas \varphi_{0}(\xv)^{*} \phi(\xv)$.
In either case the mean field undergoes visible change of shape but nonetheless most of the population remains in the ground state (c,d). Note the different time-scales for the two scenarios, and that the spatial axis is a \emph{radial} coordinate. 
\label{condevol1}}
\end{figure}

We consider two cases: scenario I with $\sub{a}{dyn}=-12a_{0}$ and scenario II with $\sub{a}{dyn}=+12a_{0}$. 
In scenario I the condensate contracts, as expected. We evolve it for $5$~ms, which is less than the Gross-Pitaevskii/HFB collapse time \cite{jila:nova} of about $7.5$~ms for this interaction strength \cite{wuester:nova}. More than 96\% of the population stays in the trap ground state mode for these $5$~ms. All this is shown in \fref{condevol1}. For the repulsive interactions in scenario II, the BEC is stable and we evolve it for the longer span of $45$~ms. During this time the cloud performs roughly one breathing oscillation as shown in \frefp{condevol1}{b}. Less population is transferred to non-ground state modes than in the attractive case.

\begin{figure}
\centering
\epsfig{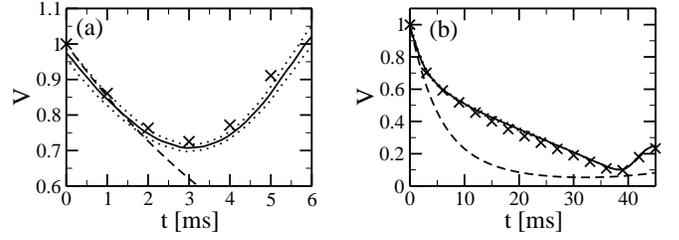} 
\caption{Minimum quadrature variances $V=\mbox{min}_{\theta}[\Delta \hat{X}^{\theta}]^2$ corresponding to maximal squeezing versus evolution time. We compare HFB simulations ($\times$, \eref{grdstuncert}) with truncated Wigner results (solid line, \eref{quadrat_uncert_twa}, the dotted lines indicate the sampling error). (a) Scenario I. (b) Scenario II. Both panels also include the corresponding analytical result for the single mode Kerr effect (dashed). We use Eq.~(1) of \rref{matthias:simon:kerr}, with $\chi=0.0134$.
\label{kerr_squeezing}}
\end{figure}

We determine the evolution of the quantum state of the condensate in the HFB approximation. The initial state is a coherent state, with $G_{A}=G_{N}=0$, and these correlation functions evolve nonzero values describing the Kerr squeezing. Our numerical results are shown in \fref{kerr_squeezing}.
They show significant quadrature squeezing, which evolves to a maximum and then decreases. We find the largest squeezing in the repulsive case (Scenario II): up to $9$ dB. These results are compared with the predictions of the single mode model of \sref{oscimodes}. For the repulsive scenario II the single mode model approximates the maximal squeezing attained and the time scale on which it occurs. For the attractive scenario I the single mode model is accurate only for short times. This is due to the contractive dynamical instability resulting in greater production of uncondensed atoms than in the repulsive case \cite{wuester:nova2,beatka:train}. 

\section{Matter-wave homodyne scheme \label{homodyne}}

Experimentally one can measure the uncertainty of observables, such as the total atom number in a BEC, by determining the shot-to-shot variance. 
However, the Hamiltonian \bref{Hamiltonian} commutes with the number operator $\hat{N}=\intas \hat{\Psi}^{\dagger}\hat{\Psi}$, so it, and its statistics, are conserved. The squeezing described in \sref{groundstate} must therefore occur along some quadrature angle $\theta\neq0$ of \eref{thetaquadratops} and cannot be experimentally measured without the phase-reference provided by the local oscillator of a homodyne scheme \cite{book:walls:milburn}. Homodyne detection is well established in quantum optics. Its core ingredient is a strong laser beam whose quantum state is coherent. Its phase provides the reference necessary to extract the noise amplitude in any quadrature. In this section we propose a homodyne scheme for Bose-Einstein condensates in a harmonic trap, making use of interference between different hyperfine states of the condensed atoms.

To this end, we consider a condensate with two components denoted by $|1\rangle$ and $|2\rangle$. Atoms are converted between components by applying electromagnetic fields; for example using microwave and RF fields in the $^{87}$Rb experiments of \rref{hall:scattlength,hall:phase}. The Hamiltonian for our two component Bose gas is hence
\begin{align}
\hat{H}&=\intas \Big\{
\sum_{i=1,2}\hat{\Psi}^{\dagger}_{i}\left(-\frac{\hbar^{2} \nabla^{2}}{2m} + V  \right)\hat{\Psi}_{i} 
\CR
&+ \sum_{i,j=1,2}\frac{U_{ij}}{2}\hat{\Psi}^{\dagger}_{i}\hat{\Psi}^{\dagger}_{j}\hat{\Psi}_{j}\hat{\Psi}_{i}
+\Omega \hat{\Psi}^{\dagger}_{1}\hat{\Psi}_{2} + \Omega^{*} \hat{\Psi}^{\dagger}_{2}\hat{\Psi}_{1}
\Big\}.
\label{two_comp_hamil}
\end{align}
The field operator $\Psi_{i}(\xv)$ carries a hyperfine index $i=1,2$ and its spatial argument is suppressed in \eref{two_comp_hamil}. We assume identical traps for both components. There are two intraspecies ($U_{11}$, $U_{22}$) and one interspecies ($U_{12}=U_{21}$) interaction strength. The magnitude of the coupling coefficient  $|\Omega|$, the Rabi frequency, controls the rate of conversion between the species. The phase of $\Omega$ controls the relative phase of the two atomic components (see Sec. \ref{mixing}). 

%

\subsection{Homodyne detection \label{detection}}

Let $\hat{\Psi}$ denote an atomic field, represented by component $|1\rangle$, that undergoes Kerr squeezing. The spatial argument $\xv$ is suppressed and we will denote $\hat{\Psi}(\xdv)$ by $\hat{\Psi}'$.
Let $\hat{\Phi}$ denote the local oscillator atomic field, represented by component $|2\rangle$,  which we assume to be at all times in a coherent state $\expec{\hat{\Phi}(\xv)}=b(\xv)e^{i\theta(\xv)}$, with $b(\xv)$ and $\theta(\xv)$ some real functions. We also assume a large amplitude for the local oscillator,
\begin{align}
|b(x)|^{2}&\gg \expec{\hat{\Psi}^{\dagger}(\xv)\hat{\Psi}(\xv)}.
\label{strong_lo}
\end{align}
In the following we denote $\hat{X}^{\theta}=\hat{\Psi}e^{-i\theta} +\hat{\Psi}^{\dagger}e^{i\theta}$, where $\theta (\xv)$ is abbreviated to $\theta$, suppressing the spatial dependence, and $\theta (\xv')$ is abbreviated to $\theta'$.

We denote the combined field $\hat{\varphi}=\hat{\Psi}+\hat{\Phi}$, and determine its number variance;  ~$[\Delta \sub{\hat{N}}{tot}]^{2}=\expec{\sub{\hat{N}}{tot}^{2}}-\expec{\sub{\hat{N}}{tot}}^{2}$ for $\sub{\hat{N}}{tot}=
\intas\hat{\varphi}^{\dagger}(\xv)\hat{\varphi}(\xv)$. The total number uncertainty can be obtained from:
\begin{align}
&\expec{\sub{\hat{N}}{tot}^{2}}=\intas\intass
\CR
& \times\expec{\left(\hat{\Psi}+\hat{\Phi}\right)^{\dagger}\left( \hat{\Psi}+\hat{\Phi}\right)\left( \hat{\Psi}'+\hat{\Phi}'\right)^{\dagger}\left( \hat{\Psi}'+\hat{\Phi}'\right)}
\CR
&=\intas\intass\Big\{\expec{\hat{\Psi}^{\dagger}\hat{\Psi}\hat{\Psi}^{\dagger'}\hat{\Psi}'}
+\expec{\hat{\Psi}^{\dagger}}b^{2'}b e^{i\theta}
\CR
&+\expec{\hat{\Psi}^{\dagger '}}b^{2}b' e^{i\theta'}
+\expec{\hat{\Psi}}b^{2'}b e^{-i\theta}+\expec{\hat{\Psi} '}b^{2}b' e^{-i\theta'}
\CR
&+b^{2}\expec{\hat{\Psi}^{\dagger '}\hat{\Psi}'} +b^{2'}\expec{\hat{\Psi}^{\dagger}\hat{\Psi}}+bb'e^{-i(\theta' -\theta)}\expec{\hat{\Psi}^{\dagger}\hat{\Psi}'}
\CR
&+bb'e^{-i(\theta -\theta')}\expec{\hat{\Psi}\hat{\Psi}^{\dagger'}}
+bb'e^{-i(\theta +\theta')}\expec{\hat{\Psi}\hat{\Psi}'}
\CR
&+bb'e^{i(\theta +\theta')}\expec{\hat{\Psi}^{\dagger '}\hat{\Psi}^{\dagger}}
+b^{2}b^{2'}+be^{-i\theta}\expec{\hat{\Psi}\hat{\Psi}^{\dagger'}\hat{\Psi}'}
\CR
&+be^{i\theta}\expec{\hat{\Psi}^{\dagger}\hat{\Psi}^{\dagger'}\hat{\Psi}'}
+b'e^{-i\theta'}\expec{\hat{\Psi}^{\dagger}\hat{\Psi}\hat{\Psi}'}
\CR
&+b'e^{i\theta'}\expec{\hat{\Psi}^{\dagger}\hat{\Psi}\hat{\Psi}^{\dagger'}} ,
\label{term1}
\end{align}
and
\begin{align}
&\expec{\sub{\hat{N}}{tot}}
=\intas\left(\expec{\hat{\Psi}^{\dagger}\hat{\Psi}} +b(\xv)\expec{\hat{X}^{\theta}}+b(\xv)^{2} \right) .
\label{term2}
\end{align}
In writing \eref{term1} we have factored the correlations between the fields $\hat{\Psi}$ and $\hat{\Phi}$. With a strong local oscillator, \eref{strong_lo}, we need only retain the leading order in $b$, and obtain:
\begin{align}
&[\Delta \sub{\hat{N}}{tot}]^{2}=\intas\intass b(\xv)b(\xdv)
\CR
&\times\Big\{
e^{-i(\theta - \theta')}\left[\delta^{3}(\xv -\xdv) +G_{N}(\xv,\xdv) \right]
+e^{i (\theta- \theta')}G_{N}(\xdv,\xv) 
\CR
&+ e^{-i(\theta+ \theta')}G_{A}(\xv,\xdv)
+ e^{i(\theta+ \theta')}G_{A}^{*}(\xv,\xdv)
\Big\}
\CR
&=\sub{N}{tot}\Big( 
1 + 2\intas\intass \varphi_{0}(\xv)\varphi_{0}(\xdv)  \bigg[  e^{-i(\theta' -\theta)} G_{N}(\xv,\xdv)
\CR
&+ \mathfrak{Re}\left\{ \intas\intass e^{-i(\theta' + \theta)} G_{A}(\xv,\xdv)
\right\}
 \bigg]  \Big).
\label{numbervariancemixed}
\end{align}
For the last equality, we have assumed that the local oscillator is in the trap ground state, so that $b(\xv)=\varphi_{0}(\xv)\sqrt{N_{2}}\simeq\varphi_{0}(\xv)\sqrt{\sub{N}{tot}}$. $N_{2}$ is the atom number in the local oscillator and we have noted that the local oscillator is highly populated compared to the squeezed field. If we also assume that the phase of the local oscillator is homogenous $\theta(\xv)=\theta(\xdv)=\theta$ we can write:
\begin{align}
[\Delta \sub{\hat{N}}{tot}]^{2}&=\sub{N}{tot}[\Delta \hat{X}_{0}^{\theta}]^{2},
\label{final_homodyne}
\end{align}
using \eref{quadrat_variance_groundst}. The quadrature angle $\theta$ is here given by the phase angle of the local oscillator condensate. Through the mixing of the local oscillator with the squeezed field, the quadrature of reduced uncertainty can be rotated into the number ``quadrature''.

\subsection{Component mixing \label{mixing}}

The experimental scheme we are modeling uses two pulses of radiation resulting in a non-zero $\Omega$ in \eref{two_comp_hamil}. The first creates a small component $| 1 \rangle$ field from an initial, large, non-interacting component $| 2 \rangle$ field. Component $| 1 \rangle$ is the field whose squeezing we seek to measure, and component $| 2 \rangle$ serves as the local oscillator. Between the first and second pulses the Kerr squeezing evolves, due to self-interaction in the $| 1 \rangle$ component. The second pulse, a time $\sub{t}{evolve}$ after the first, mixes the target field and the local oscillator so that the quadrature squeezing may be inferred from the number variance.

The particular quadrature measured is determined by the phase of the complex coupling $\Omega$. Since this is the phase of an RF field, it may be easily adjusted between the two pulses, allowing access to all quadratures.

We now consider the electromagnetic coupling in more detail. When it is switched on, the atoms undergo Rabi oscillations between the hyperfine components. These can be understood by isolating the relevant parts of the Heisenberg equations for the field operators
\begin{align}
i \hbar \dot{\hat{ \Psi}}_{1} = \Omega \hat{\Psi}_{2}, \:\:\:\:\:
i \hbar \dot{ \hat{\Psi}}_{2} = \Omega \hat{\Psi}_{1} .
\label{rabi_eom}
\end{align}
These have the solutions \cite{WilliamsHolland}
\begin{align}
\hat{ \Psi}_{1} &= \hat{ \Psi}_{1} (0) \cos ( |\Omega | t / \hbar )
-i \frac{\Omega}{| \Omega |} \hat{ \Psi}_{2} (0) \sin ( |\Omega | t / \hbar ) ,
\nonumber \\
\hat{ \Psi}_{2} &= \hat{ \Psi}_{2} (0) \cos ( |\Omega | t / \hbar )
-i \frac{| \Omega |}{\Omega} \hat{ \Psi}_{1} (0) \sin ( |\Omega | t / \hbar ) .
\label{rabi_solns}
\end{align}
After $t = t_{\pi/2} = \hbar \pi/ (4 \Omega )$ the trigonometric functions have the value $1/\sqrt{2}$ and each field has equal contributions from the initial fields, called a $\pi/2$ pulse, with a relative phase determined by the phase of the coupling coefficient $\Omega$. The modulus of $\Omega$ is chosen such that $t_{\pi/2}\ll \sub{t}{evolve}$.

\subsection{Feshbach resonances \label{feshbach}}

The situation described in \sref{detection} will be difficult to achieve. Since we require more atoms in the local oscillator component $|2\rangle$ than in the component in which we wish to measure Kerr squeezing $|1\rangle$, this configuration will give rise to Kerr squeezing in the local oscillator itself due to its self-interaction ($U_{22}$), invalidating the assumption for it to be in a coherent state. Further, there are interactions between the local oscillator and the component to be measured ($U_{12}$), which can even result in spatial phase separation \cite{wuester:skyrm}, and will also affect component one's Kerr squeezing. 

Ideally, we would like to set $U_{22}=U_{12}=0$ using a Feshbach resonance. However, this would require a rare coincidence of resonances in two different scattering channels. Hence we also consider  three other options for improving the situation with a Feshbach resonance:
(i) increasing $U_{11}$, possibly to the point that $U_{22}$ and $U_{12}$ can be neglected in comparison, (ii) turning off $U_{12}$ only, and (iii) turning off $U_{22}$ only. 

We will present detailed results for option (i) and briefly comment on our findings for the other options, including the ideal case.

\subsection{Two-component Hartee-Fock-Bogoliubov theory \label{hfb2c}}

To investigate the ideas of the previous sections in a full multi-mode quantum field simulation of the homodyne scheme, we employ HFB theory. For the two-component case  the simulation uses two condensates, $\phi_{i}=\langle \Psi_{i} \rangle$, and six correlation functions, as detailed in \aref{two_comp_hfb_appendix}. That appendix also gives our HFB equations of motion.

We simulate the evolution of squeezing in component $|1\rangle$ followed by mixing with a highly populated component $|2\rangle$. We show that this reduces the total number variance in one component below the shot-noise limit, given by the number in that component.

Since both $\hat{N}_{1}=\intas \hat{\Psi}_{1}^{\dagger}\hat{\Psi}_{1}$ and $\hat{N}_{2}=\intas \hat{\Psi}^{\dagger}_{2}\hat{\Psi}_{2}$ commute with the Hamiltonian \bref{two_comp_hamil} for $\Omega=0$, we expect the number variance of each component, \eref{single_comp_numbvervar1}, to change only during the mixing step. 

\subsection{Numerical results\label{numerical_homodyne}}

\begin{figure}
\centering
\epsfig{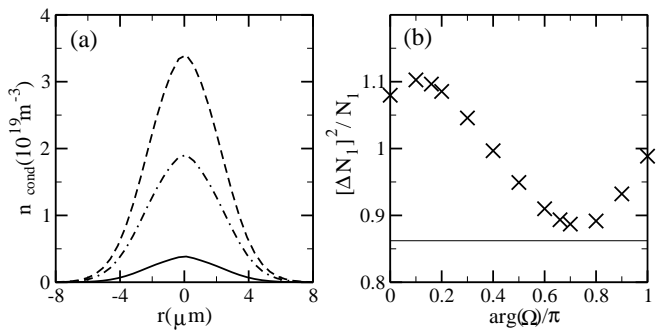} 
\caption{(a) Initial shape of ``squeezing'' condensate $|\phi_{1}(r)|^{2}$ (solid) and local oscillator $|\phi_{2}(r)|^{2}$ (dashed) and the corresponding (equal) densities after the mixing at $t=0.4$ ms (dash-dotted). (b) Number variance in condensate component $1$ after the mixing step. For $\mbox{arg}(\Omega)=0.66\pi$ the Kerr squeezing in component one with $\mbox{min}_{\theta} [X^{\theta}_{0}]^{2}=0.86$ is largely converted into detectable number squeezing. The thin line indicates the value of $0.86$ for full conversion of the quadrature squeezing.
\label{homodyne_condensate_shapes}}
\end{figure}
Now we present two-component HFB simulations of the homodyne scheme proposed in \sref{detection}. We assume the BEC is initially split into a small cloud in component $|1\rangle$, which is to be squeezed, and a larger cloud in component $|2\rangle$, to serve as the local oscillator. We envisage the following creation sequence. After condensation the BEC is adiabatically brought to a non-interacting initial state with all atoms in the same hyperfine component as done in \rref{jila:nova}. Using electromagnetically induced component mixing, the condensate is then split into small and large condensates in different states. This is the starting point of our simulations.

As discussed in \sref{feshbach}, after the splitting we assume that only \emph{one} of the three couplings ${U}_{ij}$ is tuned using a Feshbach resonance. For the other two, we used the scattering lengths of the $^{87}$Rb hyperfine components $|1\rangle=|F=1,m_{F}=-1\rangle$ and $|2\rangle=|F=2,m_{F}=1\rangle$ \cite{wuester:skyrm, cornell:scattlength}, which are $a_{11}= 100.4 a_{0}$, $a_{22}=95.47 a_{0}$ and $a_{21}=98.10 a_{0}$, where $a_{0}$ is the Bohr radius. We also considered parameters appropriate for $^{85}$Rb. For this case multi-component scattering length data is not available, thus we simply set both non-manipulated scattering lengths to $a=-443a_{0}$ \cite{jila:revision}.
The initial numbers are $N_{1}=600$, $N_{2}=5400$ both in a trap ground state. These states are shown in \frefp{homodyne_condensate_shapes}{a}. 

\begin{figure}
\centering
\epsfig{file={homodyne_U11up.eps},width=\columnwidth} 
\caption{Quantum state evolution of a two component BEC. Panels (a,c) show component $|1\rangle$, (b,d) component $|2\rangle$. The variance of the trap ground-state for the quadrature with maximal squeezing, \eref{quadrat_variance_groundst}, is shown in panels (a,b). The variance of the total number in each component, \eref{single_comp_numbvervar1},  is shown in panels (c,d) as a fraction of the total atom number in each component. For the first $0.4$ ms, $\Omega=0$, followed by a pulse of length $f \times \sub{t}{pulse}=f  \times5$ $\mu$s with $\Omega=\hbar \pi/(4 \sub{t}{pulse})$, where $f\lesssim1$ is adjusted to achieve full mixing. Subsequently, we again have $\Omega=0$. The two curves use a different reference phase: (solid) $\mbox{arg} (\Omega) =0.66\pi$, (dashed) $\mbox{arg} (\Omega) =0.16\pi$. The grayed region shows the value achieved after the component mixing. We did not evolve the system past $\sub{t}{evolve}+\sub{t}{pulse}=0.405$ ms, but continued the graph assuming constant variances until $0.6$ ms as a visualization aid.  
\label{homodyne_ideal}}
\end{figure}
For the results shown in \fref{homodyne_ideal}, we increased $a_{11}$ by a factor of five from its natural value, as in option (i). This is not enough to really make $U_{22}$ and $U_{12}$ negligible, however we found stronger interactions were not numerically tractable. In such cases we find that the local oscillator becomes squeezed as well, due to its self interaction, however $|U_{11}|> |U_{22}|$ ensures that its squeezing does not much exceed that of component one. 

For a time $\sub{t}{evolve}=0.4$ ms we see that the ground-state of component one, with $\sub{a}{dyn}= 502 a_0$, develops a minimum quadrature variance of about $0.86$ ($0.66$ dB squeezing). During this time, since the electromagnetic coupling is off, the relative number variances $[\Delta N_{i}]^{2}/N_{i}$ remain one to within the Gaussian approximation of the HFB method. The slight reduction in the relative number variances seen in \fref{homodyne_ideal}(c,d) is presumed to be due to the development of higher order correlations than can not be treated with the HFB, or TWA, method.  After $\sub{t}{evolve}$, we apply a nonzero coupling for $\sub{t}{pulse} \approx 5$ $\mu$s, until the components have mixed to equal populations, and the total number variance now reflects the quadrature variance of the ground state of component one before the pulse, for a quadrature angle $\theta$ which is controlled by the phase of the coupling, arg$(\Omega)$. Changing this phase we can pick up anti-squeezed or squeezed quadratures, shown in \frefp{homodyne_ideal}{c} and (d) and \frefp{homodyne_condensate_shapes}{b}.

We find that the homodyne scheme reduces the number fluctuations of the atom field after recombination below $N$ despite squeezing of the local oscillator. Similar findings have been reported in \cite{matthias:simon:kerr}.

In the case shown in \fref{homodyne_ideal}, corresponding to option (i) \sref{feshbach}, we find that the squeezing does not increase much with further evolution. Although, it should increase with higher values of $a_{11}$ the results presented here are sufficient to prove the principle of the homodyne scheme. For the other two options, (ii) and (iii) of \sref{feshbach}, we found that setting $U_{22}$ to zero without increasing $U_{11}$ results in an even earlier turn-around of the squeezing in component one, which therefore is negligible. Setting $U_{12}$ to zero prevents this, but without increasing $U_{11}$ the local oscillator is much more squeezed than the other component. 

Finally, we examined a $^{85}$Rb type scenario, with all three scattering lengths negative. For our simulations we increased the magnitude of $a_{11}$ by a further factor of 5. This scenario is feasible if the evolution time is much shorter than the collapse time, but shows an earlier turnaround of squeezing than the corresponding repulsive case.

\section{Conclusions \label{conclusions}}

We have shown that harmonically trapped condensates with a Feshbach resonance provide a clean and stable system to study the \emph{dynamics} of quadrature squeezing in atom-optics. We showed how to implement a matter-wave homodyne scheme and numerically demonstrated it for experimentally feasible parameters.

The insight into the squeezing evolution of vacuum-fluctuations that our scheme affords might be useful for analogue cosmology along the lines proposed in \cite{calz:hu,calz:hu2}. 

The Kerr effect studied here eventually gives rise to strongly non-Gaussian quantum fluctuations \cite{book:walls:milburn}. These have been conjectured to cause a notable discrepancy between experiment and theory in the collapse time of attractive BECs \cite{wuester:nova2}. This aspect of Kerr squeezing in attractive condensates might merit further study.

\acknowledgments
We gratefully acknowledge fruitful discussions with Graham Dennis, Jacob Dunningham, Simon Haine, Joseph Hope, Mattias Johnsson, Nick Robins and Murray Olsen. The numerical machinery for our TWA solutions was kindly provided by Matthew Davis and Blair Blakie. This research was supported by the Australian Research Council and by an award under the Merit Allocation Scheme of the National Facility of the Australian Partnership for Advanced Computing.

\appendix
\section{Quantum variances in the Hartree-Fock-Bogoliubov formalism\label{hfb_variances_appendix}}

Here we provide details regarding \eref{quadrat_variance_groundst} for a spherically symmetric situation. We decompose the oscillator eigenstates into a radial and an angular part
 $\varphi_{nlm}(\xv)=f_{nl}(r)Y_{lm}(\theta,\phi)$, where $Y_{lm}(\theta,\phi)$ is a spherical harmonic. We also expanded the collective index $k\rightarrow nlm$. Further we choose the spherical polar co-ordinate system $r'$, $\phi'$, $\theta'$ for the vector $\xdv$ such that its $z$-axis points along the vector $\xv$. $\theta'$ then denotes the angle between $\xv$ and $\xdv$, which appears as argument in the atom field correlation functions for the case of spherical symmetry \cite{wuester:nova}.
Finally we use an expansion of $G_{N}$ in terms of Legendre polynomials $P_{s}$:
\begin{align}
G_{N}(\xv,\xdv)=\sum_{s=0}^{M}G^{(s)}_{N}(r,r')P_{s}(\cos \theta'),
\label{gn_legpolexp}
\end{align}
where $r=|\xv|$, $r'=|\xdv|$. With this \eref{oscicorrfct} becomes:
\begin{align}
&\expec{\hat{a}^{\dagger}_{nlm}\hat{a}_{n'l'm'}}=\int_{0}^{\infty}\!\!\!dr r^{2}\int_{0}^{\infty}\!\!\!dr' r'^{2}
\int_{0}^{2\pi} \!\!\!d\phi \int_{-1}^{1} \!\!\!d\cos{\theta} 
\CR
& \times   \int_{0}^{2\pi} \!\!\!d\phi'  \int_{-1}^{1} \!\!\!d\cos{\theta'} 
\sum_{s=0}^{M}Y_{l'm'}(\theta',\phi')Y_{lm}^{*}(\theta,\phi)P_{s}(\cos \theta')
\CR
&\times 
G^{(s)}_{N}(r,r')f_{nl}^{*}(r)f_{n'l'}(r').
\label{oscicorrfct2}
\end{align}
For the oscillator ground state $Y_{00}=1/\sqrt{4\pi}$. This allows us to carry out the angular integrations and obtain:
\begin{align}
\expec{\hat{a}^{\dagger}_{000}\hat{a}_{000}}&=4\pi \int_{0}^{\infty} \!\!\! dr r^{2}\int_{0}^{\infty} \!\!\! dr' r'^{2}
G^{(0)}_{N}(r,r')f_{00}^{*}(r)f_{00}(r').
\label{oscicorrfct3}
\end{align}
We have used: $\int_{-1}^{1} dx P_{m}(x)= 2 \delta_{m,0}$. 
Defining $f_{00}(r)=\tilde{f}_{00}(r)/r$ and $G^{(m)}_{A/N}(r,r')=\tilde{G}^{(m)}_{A/N}(r,r')/rr'$ we finally obtain:
\begin{align}
&\mbox{min}_{\theta}\left[\Delta \hat{X}^{\theta}_{0}\right]^{2}
\CR
&=1+ 8\pi
\Big(
 \int_{0}^{\infty}dr \int_{0}^{\infty}dr'
 [\tilde{G}^{(0)}_{N}(r,r') \tilde{f}^{*}_{00}(r)\tilde{f}_{00}(r)]
\CR
&-\left|
 \int_{0}^{\infty}dr \int_{0}^{\infty}dr'
 [\tilde{G}^{(0)}_{A}(r,r') \tilde{f}_{00}(r)\tilde{f}_{00}(r)]
\right|
\Big).
\label{grdstuncert}
\end{align}
Inserting the expansion in Legendre polynomials of the correlation functions into \eref{single_comp_numbvervar1} we can also obtain:
\begin{align}
&[\Delta \hat{N}]^{2}=\sub{N}{tot} 
\CR
&+ 8\pi^{2} \int dR\int dR' \Big[ {\mathcal Re}\{\tilde\phi^{*}(R) \tilde\phi^{*}(R')
\tilde G_{A}^{(0)}(R,R') \}
\CR
&+ \tilde\phi^{*}(R) \tilde\phi(R')
\tilde G_{N}^{(0)}(R,R') \Big]
\CR
&+ 4\pi^{2}\intas\intass \sum_{s=0}^{M}\frac{1}{2s+1}
\CR
&\times\left[\left| \tilde G_{A}^{(s)}(R,R') \right|^{2} + \left| \tilde G_{N}^{(s)}(R,R') \right|^{2}  \right].
\label{single_comp_numbvervar2}
\end{align}
%

\section{Two-component Hartree-Fock-Bogoliubov equations \label{two_comp_hfb_appendix}}

To follow the quantum evolution of our two-component system through the initial squeezing stage past the mixing step, we make use of the HFB formalism.
The set of variables used in \cite{wuester:nova} must be extended to accommodate two hyperfine states. We split the field operators into mean and fluctuations:
$\hat{\Psi}_{1}=\phi_{1} +\hat{\chi}_{1}$, $\hat{\Psi}_{2}=\phi_{2} +\hat{\chi}_{2}$, with 
$\expec{\hat{\Psi}_{n}}=\phi_{n}$. We then consider two condensates $\phi_{1}$, $\phi_{2}$ and six correlation functions (using $i\in\{1,2\}$): 
\begin{align}
&G_{Ni}(\xv,\xdv)=\expec{\hat{\chi}_{i}^{\dagger}(\xdv)\hat{\chi}_{i}(\xv)},
\\
&G_{Ai}(\xv,\xdv)=\expec{\hat{\chi}_{i}(\xdv)\hat{\chi}_{i}(\xv)},
\\
&G_{CN}(\xv,\xdv)=\expec{\hat{\chi}_{2}^{\dagger}(\xdv)\hat{\chi}_{1}(\xv)},
\\
&G_{CA}(\xv,\xdv)=\expec{\hat{\chi}_{2}(\xdv)\hat{\chi}_{1}(\xv)}.
\end{align}
%

The equations of motion for the condensates and the correlation functions $G_{A/N, 1/2}$ are partially identical to those previously presented \cite{wuester:nova}. However, additional terms exist in all of them due to the coupling between the two hyperfine components. 
In the following we use the notation of \rref{wuester:nova}, in particular: $\bar{G}_{N1}\equiv \bar{G}_{N1}(x)\equiv G_{N1}(\xv,\xv)$, $\phi_{n}\equiv \phi_{n}(\xv)$, $\phi_{n}'\equiv \phi_{n}(\xdv)$ etc. We further introduce the abbreviations: $n_{j}=|\phi_{j}|^{2} + \bar{G}_{Nj}$, $\eta_{j}=\phi_{j}^{2} +\bar{G}_{Aj}$, $\xi=\phi_{2} \phi_{1} + \bar{G}_{CA}$, $\zeta=\phi_{2}^{*}\phi_{1}+\bar{G}_{CN}$ and $H_{0j}=-\hbar^{2}\nabla_{\xv}^{2}/(2m) +V(\xv) + \nu_{j}$, $H_{0j}'=-\hbar^{2}\nabla_{\xdv}^{2}/(2m) +V(\xdv) + \nu_{j}$, where we allowed possibly different detunings $\nu_{j}$ (not used in the present work).  
Our equations of motion are then:
\begin{align}
i\hbar \dot{\phi}_{1}=&H_{01}\phi_{1} +U_{11}\left(\left[2\bar{G}_{N1} + |\phi_{1}|^{2} \right]\phi_{1} +\bar{G}_{A1}\phi_{1}^{*}
\right)
\CR
&+U_{12}\big(n_{2}\phi_{1} 
+\bar{G}_{CN} \phi_{2} + \bar{G}_{CA}\phi^{*}_{2} \big) + \Omega \phi_{2},
\label{two_comp_hfb_eqns_conds}
\end{align}
\begin{align}
i\hbar \dot{G}_{N1}(\xv,\xdv)&=\left(H_{01} -H_{01}'\right)G_{N1}(\xv,\xdv)
\CR
&+U_{11}\Big\{
2\left[n_{1} - n_{1}' \right]G_{N1}(\xv,\xdv)
\CR
&+ \eta_{1}G_{A1}(\xv,\xdv)^{*}  - \eta'^{*}_{1}G_{A1}(\xv,\xdv)
\Big\}
\CR
&+U_{12}\Big\{
\left(n_{2} -n_{2}'\right)G_{N1}(\xv,\xdv) 
\CR
&+\zeta G_{CN}(\xdv,\xv)^{*} 
-\zeta'^{*}G_{CN}(\xv,\xdv) 
\CR
&+\xi G_{CA}(\xdv,\xv)^{*} 
-\xi'^{*}G_{CA}(\xv,\xdv) 
\Big\}
\CR
&+\Omega G_{CN}(\xdv,\xv)^{*} - \Omega^{*} G_{CN}(\xv,\xdv),
\label{two_comp_hfb_eqns_gn1}
\end{align}
%
%
%
%
\begin{align}
i\hbar \dot{G}_{A1}(\xv,\xdv)&=\left(H_{01}  + H_{01}'\right)G_{N1}(\xv,\xdv)
\CR
&+U_{11}\Big\{
2\left[n_{1} + n_{1}' \right]G_{A1}(\xv,\xdv)
\CR
&+ \eta_{1}G_{N1}(\xv,\xdv)^{*}  + \eta'_{1}G_{N1}(\xv,\xdv)
\CR
&+\eta_{1}\delta^{(3)}(\xv-\xdv)
\Big\}
\\
\nonumber
&+U_{12}\Big\{
\left(n_{2} +n_{2}'\right)G_{A1}(\xv,\xdv) 
\CR
&+\zeta G_{CA}(\xdv,\xv) 
+\zeta' G_{CA}(\xv,\xdv) 
\CR
&+\xi G_{CN}(\xdv,\xv)
+\xi'G_{CN}(\xv,\xdv) 
\Big\}
\CR
+&\Omega \left[G_{CA}(\xv,\xdv) +  G_{CA}(\xdv,\xv)\right],
\label{two_comp_hfb_eqns_ga1}
\end{align}
%
%
%
Throughout we have used:
\begin{align}
\expec{\hat{\chi}_{1}^{\dagger}(\xdv) \hat{\chi}_{2}(\xv)}&=G_{CN}(\xdv,\xv)^{*},
\\
\expec{\hat{\chi}_{1}^{\dagger}(\xdv)\hat{\chi}_{2}^{\dagger}(\xv)}&=G_{CA}(\xdv,\xv)^{*} .
\label{relations1}
\end{align}
One can deduce the equations for component two from those of component one by using the following symmetry relations, under exchange of particle labels $1\leftrightarrow2$:
\begin{align}
G_{CN}(\xdv,\xv)^{*}& \leftrightarrow G_{CN}(\xv,\xdv),
\label{gcn_symm}
\\
G_{CA}(\xdv,\xv)& \leftrightarrow G_{CA}(\xv,\xdv),
\label{gca_symm}
\\
\bar{G}_{CN}& \leftrightarrow \bar{G}_{CN}^{*},
\\
\bar{G}_{CA}& \leftrightarrow \bar{G}_{CA},
\\
\Omega& \leftrightarrow \Omega^{*}.
\label{exchange_relation}
\end{align}
%
%
%
The equations of motion for the cross correlation functions are:
\begin{align}
&i\hbar \dot{G}_{CN}(\xv,\xdv)=
\CR
&\left(H_{01}- H_{02}' \right)G_{CN}(\xv,\xdv)
+2\left[U_{11}n_{1} - U_{22}n_{2}' \right]G_{CN}(\xv,\xdv)
\CR
&+U_{11}\eta_{1}G_{CA}^{*}(\xv,\xdv)  -U_{22}\eta_{2}'^{*}G_{CA}(\xv,\xdv) 
\CR
&+U_{12}\Big\{
\left(n_{2}-n_{1}'\right)G_{CN}(\xv,\xdv) 
\CR
&+\xi G_{A2}(\xv,\xdv)^{*} - \xi^{'*}G_{A1}(\xv,\xdv) 
\CR
&+\zeta G_{N2}(\xv,\xdv) - \zeta^{'}G_{N1}(\xv,\xdv) \Big\}
\CR
&+\Omega \left(G_{N2}(\xv,\xdv) -  G_{N1}(\xv,\xdv)\right)
\label{two_comp_hfb_eqns_gcn},
\end{align}
and
\begin{align}
&i\hbar \dot{G}_{CA}(\xv,\xdv)=
\CR
&\left(H_{01} + H_{02}' \right)G_{CA}(\xv,\xdv)
+2\left[U_{11}n_{1} + U_{22}n_{2}' \right]G_{CA}(\xv,\xdv)
\CR
&+U_{11}\eta_{1}G_{CN}^{*}(\xv,\xdv)  +U_{22}\eta_{2}'G_{CN}(\xv,\xdv) 
\CR
&+U_{12}\Big\{
\left(n_{2}+n_{1}'\right)G_{CA}(\xv,\xdv) 
\CR
&+\xi \left[ G_{N2}(\xv,\xdv)^{*} + \delta^{(3)}(\xv-\xdv)\right] + \xi^{'}G_{N1}(\xv,\xdv) 
\CR
&+\zeta G_{A2}(\xv,\xdv) +  \zeta^{'*}G_{A1}(\xv,\xdv) \Big\}
\CR
&+\Omega G_{A2}(\xv,\xdv) +  \Omega^{*} G_{A1}(\xv,\xdv)
\label{two_comp_hfb_eqns_gca}.
\end{align}
%




\begin{thebibliography}{33}
\expandafter\ifx\csname natexlab\endcsname\relax\def\natexlab#1{#1}\fi
\expandafter\ifx\csname bibnamefont\endcsname\relax
  \def\bibnamefont#1{#1}\fi
\expandafter\ifx\csname bibfnamefont\endcsname\relax
  \def\bibfnamefont#1{#1}\fi
\expandafter\ifx\csname citenamefont\endcsname\relax
  \def\citenamefont#1{#1}\fi
\expandafter\ifx\csname url\endcsname\relax
  \def\url#1{\texttt{#1}}\fi
\expandafter\ifx\csname urlprefix\endcsname\relax\def\urlprefix{URL }\fi
\providecommand{\bibinfo}[2]{#2}
\providecommand{\eprint}[2][]{\url{#2}}

\bibitem[{\citenamefont{Johnsson and Haine}(2007)}]{matthias:simon:kerr}
\bibinfo{author}{\bibfnamefont{M.~T.} \bibnamefont{Johnsson}} \bibnamefont{and}
  \bibinfo{author}{\bibfnamefont{S.~A.} \bibnamefont{Haine}},
  \bibinfo{journal}{Phys. Rev. Lett.} \textbf{\bibinfo{volume}{99}},
  \bibinfo{pages}{010401} (\bibinfo{year}{2007}).

\bibitem[{\citenamefont{Barcel{\'o} et~al.}(2005)\citenamefont{Barcel{\'o},
  Liberati, and Visser}}]{visser:review}
\bibinfo{author}{\bibfnamefont{C.}~\bibnamefont{Barcel{\'o}}},
  \bibinfo{author}{\bibfnamefont{S.}~\bibnamefont{Liberati}}, \bibnamefont{and}
  \bibinfo{author}{\bibfnamefont{M.}~\bibnamefont{Visser}},
  \bibinfo{journal}{Living Rev. Relativity} \textbf{\bibinfo{volume}{8}},
  \bibinfo{pages}{12} (\bibinfo{year}{2005}).

\bibitem[{\citenamefont{Calzetta and Hu}(2003)}]{calz:hu}
\bibinfo{author}{\bibfnamefont{E.~A.} \bibnamefont{Calzetta}} \bibnamefont{and}
  \bibinfo{author}{\bibfnamefont{B.~L.} \bibnamefont{Hu}},
  \bibinfo{journal}{Phys. Rev. A} \textbf{\bibinfo{volume}{68}},
  \bibinfo{pages}{043625} (\bibinfo{year}{2003}).

\bibitem[{\citenamefont{Calzetta and Hu}(2005)}]{calz:hu2}
\bibinfo{author}{\bibfnamefont{E.~A.} \bibnamefont{Calzetta}} \bibnamefont{and}
  \bibinfo{author}{\bibfnamefont{B.~L.} \bibnamefont{Hu}},
  \bibinfo{journal}{Int. J. Theor. Phys.} \textbf{\bibinfo{volume}{44}},
  \bibinfo{pages}{1691} (\bibinfo{year}{2005}).

\bibitem[{\citenamefont{Barcel{\'o} et~al.}(2003)\citenamefont{Barcel{\'o},
  Liberati, and Visser}}]{barcelo:cpc}
\bibinfo{author}{\bibfnamefont{C.}~\bibnamefont{Barcel{\'o}}},
  \bibinfo{author}{\bibfnamefont{S.}~\bibnamefont{Liberati}}, \bibnamefont{and}
  \bibinfo{author}{\bibfnamefont{M.}~\bibnamefont{Visser}},
  \bibinfo{journal}{Phys. Rev. A} \textbf{\bibinfo{volume}{68}},
  \bibinfo{pages}{053613} (\bibinfo{year}{2003}).

\bibitem[{\citenamefont{Savage et~al.}(2003)\citenamefont{Savage, Robins, and
  Hope}}]{savage:coll}
\bibinfo{author}{\bibfnamefont{C.~M.} \bibnamefont{Savage}},
  \bibinfo{author}{\bibfnamefont{N.~P.} \bibnamefont{Robins}},
  \bibnamefont{and} \bibinfo{author}{\bibfnamefont{J.~J.} \bibnamefont{Hope}},
  \bibinfo{journal}{Phys. Rev. A} \textbf{\bibinfo{volume}{67}},
  \bibinfo{pages}{014304} (\bibinfo{year}{2003}).

\bibitem[{\citenamefont{W{\"u}ster
  et~al.}(2005{\natexlab{a}})\citenamefont{W{\"u}ster, Hope, and
  Savage}}]{wuester:nova}
\bibinfo{author}{\bibfnamefont{S.}~\bibnamefont{W{\"u}ster}},
  \bibinfo{author}{\bibfnamefont{J.~J.} \bibnamefont{Hope}}, \bibnamefont{and}
  \bibinfo{author}{\bibfnamefont{C.~M.} \bibnamefont{Savage}},
  \bibinfo{journal}{Phys. Rev. A} \textbf{\bibinfo{volume}{71}},
  \bibinfo{pages}{033604} (\bibinfo{year}{2005}{\natexlab{a}}).

\bibitem[{\citenamefont{W{\"u}ster et~al.}(2007)\citenamefont{W{\"u}ster,
  D{\c{a}}browska-W{\"u}ster, Bradley, Davis, Blakie, Hope, and
  Savage}}]{wuester:nova2}
\bibinfo{author}{\bibfnamefont{S.}~\bibnamefont{W{\"u}ster}},
  \bibinfo{author}{\bibfnamefont{B.~J.}
  \bibnamefont{D{\c{a}}browska-W{\"u}ster}},
  \bibinfo{author}{\bibfnamefont{A.~S.} \bibnamefont{Bradley}},
  \bibinfo{author}{\bibfnamefont{M.~J.} \bibnamefont{Davis}},
  \bibinfo{author}{\bibfnamefont{P.~B.} \bibnamefont{Blakie}},
  \bibinfo{author}{\bibfnamefont{J.~J.} \bibnamefont{Hope}}, \bibnamefont{and}
  \bibinfo{author}{\bibfnamefont{C.~M.} \bibnamefont{Savage}},
  \bibinfo{journal}{Phys. Rev. A} \textbf{\bibinfo{volume}{75}},
  \bibinfo{pages}{043611} (\bibinfo{year}{2007}).

\bibitem[{\citenamefont{Weinfurtner et~al.}(2007)\citenamefont{Weinfurtner,
  White, and Visser}}]{silke:sigchange}
\bibinfo{author}{\bibfnamefont{S.}~\bibnamefont{Weinfurtner}},
  \bibinfo{author}{\bibfnamefont{A.}~\bibnamefont{White}}, \bibnamefont{and}
  \bibinfo{author}{\bibfnamefont{M.}~\bibnamefont{Visser}}
  (\bibinfo{year}{2007}), \bibinfo{note}{{Phys.~Rev.~D, in press,
  gr-qc/0703117.}}

\bibitem[{\citenamefont{Griffin}(1996)}]{griffin:gappy}
\bibinfo{author}{\bibfnamefont{A.}~\bibnamefont{Griffin}},
  \bibinfo{journal}{Phys. Rev. B} \textbf{\bibinfo{volume}{53}},
  \bibinfo{pages}{9341} (\bibinfo{year}{1996}).

\bibitem[{\citenamefont{Hutchinson et~al.}(1998)\citenamefont{Hutchinson, Dodd,
  and Burnett}}]{hutchinson:gaplessmodes}
\bibinfo{author}{\bibfnamefont{D.~A.~W.} \bibnamefont{Hutchinson}},
  \bibinfo{author}{\bibfnamefont{R.~J.} \bibnamefont{Dodd}}, \bibnamefont{and}
  \bibinfo{author}{\bibfnamefont{K.}~\bibnamefont{Burnett}},
  \bibinfo{journal}{Phys. Rev. Lett.} \textbf{\bibinfo{volume}{81}},
  \bibinfo{pages}{2198} (\bibinfo{year}{1998}).

\bibitem[{\citenamefont{Morgan}(2005)}]{morgan:thermalqft}
\bibinfo{author}{\bibfnamefont{S.~A.} \bibnamefont{Morgan}},
  \bibinfo{journal}{Phys. Rev. A} \textbf{\bibinfo{volume}{72}},
  \bibinfo{pages}{043609} (\bibinfo{year}{2005}).

\bibitem[{\citenamefont{Milstein et~al.}(2003)\citenamefont{Milstein, Menotti,
  and Holland}}]{holland:burst}
\bibinfo{author}{\bibfnamefont{J.~N.} \bibnamefont{Milstein}},
  \bibinfo{author}{\bibfnamefont{C.}~\bibnamefont{Menotti}}, \bibnamefont{and}
  \bibinfo{author}{\bibfnamefont{M.~J.} \bibnamefont{Holland}},
  \bibinfo{journal}{New J. Phys.} \textbf{\bibinfo{volume}{5}},
  \bibinfo{pages}{52} (\bibinfo{year}{2003}).

\bibitem[{\citenamefont{Steel et~al.}(1998)\citenamefont{Steel, Olsen, Plimak,
  Drummond, Tan, Collet, Walls, and Graham}}]{steel:wigner}
\bibinfo{author}{\bibfnamefont{M.~J.} \bibnamefont{Steel}},
  \bibinfo{author}{\bibfnamefont{M.~K.} \bibnamefont{Olsen}},
  \bibinfo{author}{\bibfnamefont{L.~I.} \bibnamefont{Plimak}},
  \bibinfo{author}{\bibfnamefont{P.~D.} \bibnamefont{Drummond}},
  \bibinfo{author}{\bibfnamefont{S.~M.} \bibnamefont{Tan}},
  \bibinfo{author}{\bibfnamefont{M.~J.} \bibnamefont{Collett}},
  \bibinfo{author}{\bibfnamefont{D.~F.} \bibnamefont{Walls}}, \bibnamefont{and}
  \bibinfo{author}{\bibfnamefont{R.}~\bibnamefont{Graham}},
  \bibinfo{journal}{Phys. Rev. A} \textbf{\bibinfo{volume}{58}},
  \bibinfo{pages}{4824} (\bibinfo{year}{1998}).

\bibitem[{\citenamefont{Gardiner and Zoller}(2004)}]{book:qn}
\bibinfo{author}{\bibfnamefont{C.~W.} \bibnamefont{Gardiner}} \bibnamefont{and}
  \bibinfo{author}{\bibfnamefont{P.}~\bibnamefont{Zoller}},
  \emph{\bibinfo{title}{Quantum Noise}} (\bibinfo{publisher}{Springer-Verlag,
  Berlin Heidelberg,}, \bibinfo{year}{2004}).

\bibitem[{\citenamefont{Norrie et~al.}(2005)\citenamefont{Norrie, Ballagh, and
  Gardiner}}]{norrie:prl}
\bibinfo{author}{\bibfnamefont{A.~A.} \bibnamefont{Norrie}},
  \bibinfo{author}{\bibfnamefont{R.~J.} \bibnamefont{Ballagh}},
  \bibnamefont{and} \bibinfo{author}{\bibfnamefont{C.~W.}
  \bibnamefont{Gardiner}}, \bibinfo{journal}{Phys. Rev. Lett.}
  \textbf{\bibinfo{volume}{94}}, \bibinfo{pages}{040401}
  (\bibinfo{year}{2005}).

\bibitem[{\citenamefont{Sinatra et~al.}(2002)\citenamefont{Sinatra, Lobo, and
  Castin}}]{castin:validity}
\bibinfo{author}{\bibfnamefont{A.}~\bibnamefont{Sinatra}},
  \bibinfo{author}{\bibfnamefont{C.}~\bibnamefont{Lobo}}, \bibnamefont{and}
  \bibinfo{author}{\bibfnamefont{Y.}~\bibnamefont{Castin}},
  \bibinfo{journal}{J. Phys. B: At. Mol. Opt. Phys.}
  \textbf{\bibinfo{volume}{35}}, \bibinfo{pages}{3599} (\bibinfo{year}{2002}).

\bibitem[{\citenamefont{Blakie and Davis}(2005)}]{matthewandblair:pgpe}
\bibinfo{author}{\bibfnamefont{P.~B.} \bibnamefont{Blakie}} \bibnamefont{and}
  \bibinfo{author}{\bibfnamefont{M.~J.} \bibnamefont{Davis}},
  \bibinfo{journal}{Phys. Rev. A} \textbf{\bibinfo{volume}{72}},
  \bibinfo{pages}{063608} (\bibinfo{year}{2005}).

\bibitem[{\citenamefont{Norrie}(2005)}]{norrie:thesis}
\bibinfo{author}{\bibfnamefont{A.~A.} \bibnamefont{Norrie}}, Ph.D. thesis,
  \bibinfo{school}{University of Otago} (\bibinfo{year}{2005}),
  \urlprefix\url{http://www.physics.otago.ac.nz/research/jackdodd/resources/th%
esis_page.html}.

\bibitem[{\citenamefont{Dunningham et~al.}(1998)\citenamefont{Dunningham,
  Collet, and Walls}}]{dunningham:sqz}
\bibinfo{author}{\bibfnamefont{J.~A.} \bibnamefont{Dunningham}},
  \bibinfo{author}{\bibfnamefont{M.~J.} \bibnamefont{Collet}},
  \bibnamefont{and} \bibinfo{author}{\bibfnamefont{D.~F.} \bibnamefont{Walls}},
  \bibinfo{journal}{Physics Letters A} \textbf{\bibinfo{volume}{245}},
  \bibinfo{pages}{49} (\bibinfo{year}{1998}).

\bibitem[{\citenamefont{Lewenstein and You}(1996)}]{lewenstein:phasediff}
\bibinfo{author}{\bibfnamefont{M.}~\bibnamefont{Lewenstein}} \bibnamefont{and}
  \bibinfo{author}{\bibfnamefont{L.}~\bibnamefont{You}},
  \bibinfo{journal}{Phys. Rev. Lett.} \textbf{\bibinfo{volume}{77}},
  \bibinfo{pages}{3489} (\bibinfo{year}{1996}).

\bibitem[{\citenamefont{Haque and Ruckenstein}(2006)}]{haque:squeezing}
\bibinfo{author}{\bibfnamefont{M.}~\bibnamefont{Haque}} \bibnamefont{and}
  \bibinfo{author}{\bibfnamefont{A.~E.} \bibnamefont{Ruckenstein}},
  \bibinfo{journal}{Phys. Rev. A} \textbf{\bibinfo{volume}{74}},
  \bibinfo{pages}{043622} (\bibinfo{year}{2006}).

\bibitem[{\citenamefont{M{\o}lmer}(1997)}]{molmer:cohstate}
\bibinfo{author}{\bibfnamefont{K.}~\bibnamefont{M{\o}lmer}},
  \bibinfo{journal}{Phys. Rev. A} \textbf{\bibinfo{volume}{55}},
  \bibinfo{pages}{3195} (\bibinfo{year}{1997}).

\bibitem[{\citenamefont{Donley et~al.}(2001)\citenamefont{Donley, Claussen,
  Cornish, Roberts, Cornell, and Wieman}}]{jila:nova}
\bibinfo{author}{\bibfnamefont{E.~A.} \bibnamefont{Donley}},
  \bibinfo{author}{\bibfnamefont{N.~R.} \bibnamefont{Claussen}},
  \bibinfo{author}{\bibfnamefont{S.~L.} \bibnamefont{Cornish}},
  \bibinfo{author}{\bibfnamefont{J.~L.} \bibnamefont{Roberts}},
  \bibinfo{author}{\bibfnamefont{E.~A.} \bibnamefont{Cornell}},
  \bibnamefont{and} \bibinfo{author}{\bibfnamefont{C.~E.}
  \bibnamefont{Wieman}}, \bibinfo{journal}{Nature}
  \textbf{\bibinfo{volume}{412}}, \bibinfo{pages}{295} (\bibinfo{year}{2001}).

\bibitem[{\citenamefont{Walls and Milburn}(1994)}]{book:walls:milburn}
\bibinfo{author}{\bibfnamefont{D.~F.} \bibnamefont{Walls}} \bibnamefont{and}
  \bibinfo{author}{\bibfnamefont{G.~J.} \bibnamefont{Milburn}},
  \emph{\bibinfo{title}{Quantum Optics}} (\bibinfo{publisher}{Springer Verlag},
  \bibinfo{year}{1994}).

\bibitem[{\citenamefont{Kokkelmans and Holland}(2002)}]{holland:deltamu}
\bibinfo{author}{\bibfnamefont{S.~J. J. M.~F.} \bibnamefont{Kokkelmans}}
  \bibnamefont{and} \bibinfo{author}{\bibfnamefont{M.~J.}
  \bibnamefont{Holland}}, \bibinfo{journal}{Phys. Rev. Lett.}
  \textbf{\bibinfo{volume}{89}}, \bibinfo{pages}{180401}
  (\bibinfo{year}{2002}).

\bibitem[{\citenamefont{D\c{a}browska-W{\"u}ster
  et~al.}(2006)\citenamefont{D\c{a}browska-W{\"u}ster, W{\"u}ster, Bradley,
  Davis, and Ostrovskaya}}]{beatka:train}
\bibinfo{author}{\bibfnamefont{B.~J.} \bibnamefont{D\c{a}browska-W{\"u}ster}},
  \bibinfo{author}{\bibfnamefont{S.}~\bibnamefont{W{\"u}ster}},
  \bibinfo{author}{\bibfnamefont{A.~S.} \bibnamefont{Bradley}},
  \bibinfo{author}{\bibfnamefont{M.~J.} \bibnamefont{Davis}}, \bibnamefont{and}
  \bibinfo{author}{\bibfnamefont{E.~A.} \bibnamefont{Ostrovskaya}}
  (\bibinfo{year}{2006}), \eprint{cond-mat/0607332}.

\bibitem[{\citenamefont{Hall et~al.}(1998{\natexlab{a}})\citenamefont{Hall,
  Matthews, Ensher, Wieman, and Cornell}}]{hall:scattlength}
\bibinfo{author}{\bibfnamefont{D.~S.} \bibnamefont{Hall}},
  \bibinfo{author}{\bibfnamefont{M.~R.} \bibnamefont{Matthews}},
  \bibinfo{author}{\bibfnamefont{J.~R.} \bibnamefont{Ensher}},
  \bibinfo{author}{\bibfnamefont{C.~E.} \bibnamefont{Wieman}},
  \bibnamefont{and} \bibinfo{author}{\bibfnamefont{E.~A.}
  \bibnamefont{Cornell}}, \bibinfo{journal}{Phys. Rev. Lett.}
  \textbf{\bibinfo{volume}{81}}, \bibinfo{pages}{1539}
  (\bibinfo{year}{1998}{\natexlab{a}}).

\bibitem[{\citenamefont{Hall et~al.}(1998{\natexlab{b}})\citenamefont{Hall,
  M.~R.~Matthews, and Cornell}}]{hall:phase}
\bibinfo{author}{\bibfnamefont{D.~S.} \bibnamefont{Hall}},
\bibinfo{author} {\bibnamefont{M.~R.} \bibnamefont{Matthews}},
\bibinfo{author}{\bibfnamefont{C.~E.}  \bibnamefont{Wieman}},
  \bibnamefont{and} \bibinfo{author}{\bibfnamefont{E.~A.}
  \bibnamefont{Cornell}}, \bibinfo{journal}{Phys. Rev. Lett.}
  \textbf{\bibinfo{volume}{81}}, \bibinfo{pages}{1543}
  (\bibinfo{year}{1998}{\natexlab{b}}).

\bibitem[{\citenamefont{Williams and Holland}(1999)}]{WilliamsHolland}
\bibinfo{author}{\bibfnamefont{J.~E.} \bibnamefont{Williams}} \bibnamefont{and}
  \bibinfo{author}{\bibfnamefont{M.~J.} \bibnamefont{Holland}},
  \bibinfo{journal}{Nature} \textbf{\bibinfo{volume}{401}},
  \bibinfo{pages}{568} (\bibinfo{year}{1999}).

\bibitem[{\citenamefont{W{\"u}ster
  et~al.}(2005{\natexlab{b}})\citenamefont{W{\"u}ster, Argue, and
  Savage}}]{wuester:skyrm}
\bibinfo{author}{\bibfnamefont{S.}~\bibnamefont{W{\"u}ster}},
  \bibinfo{author}{\bibfnamefont{T.~E.} \bibnamefont{Argue}}, \bibnamefont{and}
  \bibinfo{author}{\bibfnamefont{C.~M.} \bibnamefont{Savage}},
  \bibinfo{journal}{Phys. Rev. A} \textbf{\bibinfo{volume}{72}},
  \bibinfo{pages}{043616} (\bibinfo{year}{2005}{\natexlab{b}}).

\bibitem[{\citenamefont{Harber et~al.}(2002)\citenamefont{Harber, Lewandowski,
  McGuirk, and Cornell}}]{cornell:scattlength}
\bibinfo{author}{\bibfnamefont{D.~M.} \bibnamefont{Harber}},
  \bibinfo{author}{\bibfnamefont{H.~J.} \bibnamefont{Lewandowski}},
  \bibinfo{author}{\bibfnamefont{J.~M.} \bibnamefont{McGuirk}},
  \bibnamefont{and} \bibinfo{author}{\bibfnamefont{E.~A.}
  \bibnamefont{Cornell}}, \bibinfo{journal}{Phys. Rev. A}
  \textbf{\bibinfo{volume}{66}}, \bibinfo{pages}{053616}
  (\bibinfo{year}{2002}).

\bibitem[{\citenamefont{Claussen et~al.}(2003)\citenamefont{Claussen,
  Kokkelmans, Thompson, Donley, Hodby, and Wieman}}]{jila:revision}
\bibinfo{author}{\bibfnamefont{N.~R.} \bibnamefont{Claussen}},
  \bibinfo{author}{\bibfnamefont{S.~J. J. M.~F.} \bibnamefont{Kokkelmans}},
  \bibinfo{author}{\bibfnamefont{S.~T.} \bibnamefont{Thompson}},
  \bibinfo{author}{\bibfnamefont{E.~A.} \bibnamefont{Donley}},
  \bibinfo{author}{\bibfnamefont{E.}~\bibnamefont{Hodby}}, \bibnamefont{and}
  \bibinfo{author}{\bibfnamefont{C.~E.} \bibnamefont{Wieman}},
  \bibinfo{journal}{Phys. Rev. A} \textbf{\bibinfo{volume}{67}},
  \bibinfo{pages}{060701(R)} (\bibinfo{year}{2003}).

\end{thebibliography}
\end{document}